\documentclass[12pt,reprint,prd,twocolumn]{revtex4}
\usepackage{amsmath,amssymb,epsfig,
latexsym,axodraw}
\usepackage{graphicx,epsfig}
\usepackage{times}
\usepackage{amsmath}

\newcommand{\amuSUOL}{a_\mu^{\rm 1L}}
\newcommand{\amuFSf}{a_\mu^{{\rm 2L,} f\tilde{f}}}
\begin{document}
\begin{flushright}
\end{flushright}
\title{\Large\bf 
Non-decoupling two-loop corrections to
 \boldmath{$(g-2)_\mu$} \\ from fermion/sfermion loops in the MSSM}
\author{H.\ G.\ Fargnoli$^{a,b}$, C.\ Gnendiger$^a$, \\ S.\ Pa{\normalfont\ss}ehr$^c$, D.\ St\"ockinger$^a$,
H.\ St\"ockinger-Kim$^a$ \vspace{1em}
}
\affiliation{\it ${}^a$Institut f\"ur Kern- und Teilchenphysik,
TU Dresden, Dresden, Germany}
\affiliation{\it ${}^b$Universidade Federal de Lavras, Lavras, Brazil}
\affiliation{\it ${}^c$Max-Planck Institut f\"ur Physik,
M\"unchen, Germany}
\setcounter{footnote}{0}
\begin{abstract}
Two-loop contributions to the muon $(g-2)$ from fermion/sfermion
loops in the MSSM are presented, and an overview of the full MSSM
prediction for $(g-2)$ is given with emphasis on the behaviour in
scenarios which are compatible with LHC data, including scenarios with
large mass splittings. Compared to all previously known 
two-loop contributions, the fermion/sfermion-loop contributions can yield the
largest numerical results.  
The new contributions contain the important universal quantities
$\Delta\alpha$ and $\Delta\rho$, and for large sfermion masses the
contributions are non-decoupling and logarithmically enhanced. We find up to $15\%$ ($30\%$)
corrections for sfermion masses in the 20 TeV (1000 TeV) range.
\end{abstract}

\maketitle 

\section{Introduction}

Supersymmetry (SUSY) is one of the best motivated ideas for physics
beyond the Standard Model (SM) at the TeV scale. 
%
The anomalous magnetic moment of the muon $a_\mu=(g-2)_\mu/2$ provides
an important constraint on physics beyond the SM and a hint for
low-energy SUSY. In the
last decade, a discrepancy between the experimental value \cite{Bennett:2006} and the SM
prediction has continuously consolidated. It now amounts
to 
\begin{align}
a_\mu^{\rm Exp-SM}&= (28.7 \pm 8.0 )  \times 10^{-10}
\label{deviation}
\end{align}
according to Ref.\ \cite{Gnendiger:2013pva}, based on the hadronic
contributions of Ref.\ \cite{Davier}. For
further recent  evaluations and reviews, see Refs.\
\cite{HMNT,Benayoun:2012wc,MdRRS}.


The deviation (\ref{deviation}) can be explained very well by
the extra contributions from low-energy SUSY in the minimal
supersymmetric standard model (MSSM). However, a tension between
$a_\mu$, which prefers light SUSY particles, and LHC-results, which
rule out certain MSSM parameter regions with light coloured SUSY
particles, seems to develop. In the
general MSSM there is no problem to accommodate all constraints
simultaneously \cite{Benbrik:2012rm,Arbey:2012dq}, but in more specific models, such as the
Constrained MSSM \cite{Bechtle:2012zk,Balazs:2012qc,Buchmueller:2012hv} or the scenarios of Refs.\
\cite{Baer:2012uy,Papucci:2011wy} motivated by the little hierarchy
problem, it has become impossible to explain the deviation
(\ref{deviation}).  Hence the new experimental input has motivated the
construction and analysis of many new SUSY models and scenarios. Some
involve compressed spectra \cite{LeCompte:2011fh,Murayama:2012jh} to
allow lighter coloured SUSY particles without contradicting LHC
limits. Some go beyond the MSSM and reconcile the Higgs mass with
$a_\mu$ by  lifting the Higgs mass with extra matter
\cite{Endo:2012cc,Endo:2011mc,Moroi:2011aa} 
or gauge bosons \cite{Endo:2011gy}. 
Many recently proposed models which stay within the MSSM framework
involve rather split spectra, e.g.\ heavy coloured, light non-coloured 
SUSY particles \cite{Endo:2013bba,Ibe:2012qu,1304.2508,Cheng:2013hna}, heavy third family,
lighter first and second family \cite{Ibe:2013oha}, non-universal
gaugino masses \cite{Mohanty:2013soa,Akula:2013ioa}, large Higgsino
masses and large stop mixing from more generic gauge mediation
\cite{Evans:2012hg}.


Not only the LHC data but also the $a_\mu$ measurement will improve
soon. The Fermilab $g-2$ experiment \cite{Carey:2009zzb,Roberts:2010cj} is under
construction. It aims to reduce the uncertainty
by more than a factor four, down to $1.6\times10^{-10}$. At
J-PARC~\cite{Iinuma:2011zz} a complementary $g-2$ experiment with similar
precision goal is planned. Clearly, pinning down the deviation
(\ref{deviation}) more precisely will lead to very valuable
constraints on SUSY which are complementary to the ones from LHC data
\cite{WhitePaper,MdRRS}. 


The prospective experimental precision calls for similar improvements
on the theory side, not only of the SM but also of the SUSY prediction
for $a_\mu$. The present theory error of the SUSY prediction has been
estimated to $3\times10^{-10}$ \cite{DSreview}, almost twice as large
as the future experimental uncertainty.  


\begin{figure*}
\begin{center}
\null\hfill
\scalebox{.65}{\setlength{\unitlength}{1pt}
\begin{picture}(240,100)(-120,0)
\CArc(0,0)(60,0,180)
\ArrowLine(-120,0)(-60,0)
\ArrowLine(60,0)(120,0)
\DashArrowLine(-60,0)(60,0){4}
\Photon(-100,100)(-60,60){4}{5}
\Vertex(60,0){2}
\Vertex(-60,0){2}
\Text(-80,95)[]{\scalebox{1.35}{$\gamma$}}
\Text(-100,-10)[]{\scalebox{1.35}{$\mu$}}
\Text(100,-10)[]{\scalebox{1.35}{${\mu}$}}
\Text(0,-10)[c]{\scalebox{1.35}{$\tilde{\mu}, \tilde{\nu}_{\mu}$}}
\Text(0,50)[c]{\scalebox{1.35}{$\tilde{\chi} _{i} ^{0,-}$}}
\end{picture}}
\hfill
\scalebox{.65}{\setlength{\unitlength}{1pt}
\begin{picture}(240,100)(-120,0)
\CArc(0,0)(60,0,60)
\CArc(0,0)(60,120,180)
\ArrowArcn(0,51)(30,180,0)
\DashArrowArcn(0,51)(30,0,180){4}
\ArrowLine(-120,0)(-60,0)
\ArrowLine(60,0)(120,0)
\DashArrowLine(-60,0)(60,0){4}
\Photon(-100,100)(-60,60){4}{5}
\Vertex(30,51){2}
\Vertex(-30,51){2}
\Vertex(60,0){2}
\Vertex(-60,0){2}
\Text(-80,95)[]{\scalebox{1.35}{$\gamma$}}
\Text(-100,-10)[]{\scalebox{1.35}{$\mu$}}
\Text(100,-10)[]{\scalebox{1.35}{${\mu}$}}
\Text(0,-10)[c]{\scalebox{1.35}{$\tilde{\mu}, \tilde{\nu}_{\mu}$}}
\Text(-40,20)[]{\scalebox{1.35}{$\tilde{\chi} _{j} ^{0,-}$}}
\Text(40,20)[]{\scalebox{1.35}{$\tilde{\chi} _{i} ^{0,-}$}}
\Text(0,70)[]{\scalebox{1.35}{$f,f'$}}
\Text(0,32)[]{\scalebox{1.35}{$\tilde f$}}
\end{picture}}
\hfill\null
\end{center}
\caption{\label{fig:diagrams}
SUSY one-loop diagrams and two-loop diagrams with
  fermion/sfermion-loop insertion. The
  photon can couple to each charged particle.}
\end{figure*}
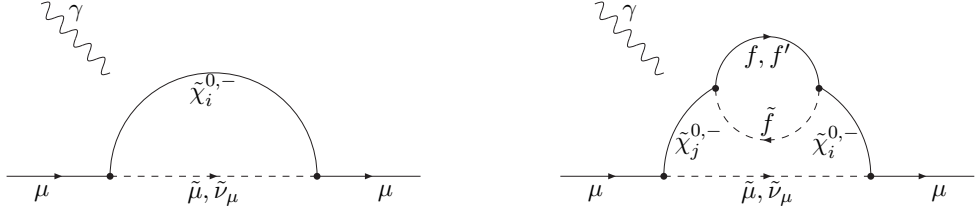

In this work we present results of a new calculation of an important
class of SUSY two-loop contributions to $a_\mu$: contributions where a
fermion/sfermion loop is inserted into a SUSY one-loop diagram. Fig.\
\ref{fig:diagrams} shows the generic diagram of this class,
along with the SUSY one-loop diagrams. The computation of these
diagrams is technically more demanding than the ones of the previously known
two-loop contributions to $a_\mu$. However, these two-loop
fermion/sfermion-loop 
contributions are phenomenologically interesting particularly for
split spectra because they are non-decoupling but logarithmically
enhanced by heavy squarks. They further eliminate a significant
source of theory uncertainty present in the one-loop
contributions.

In the following we begin with an overview of the known contributions,
stressing the variety of possibilities for large SUSY contributions to
$a_\mu$. We define benchmark points representing these
possibilities. We then explain the new contributions from
fermion/sfermion loops and their numerical behaviour. Finally we give
a comparison of the
new and the previously known contributions.

\section{Overview of SUSY contributions to \boldmath{$\lowercase{a}_\mu$}}

\begin{figure*}
\begin{center}
\epsfxsize=0.5\textwidth\epsfbox{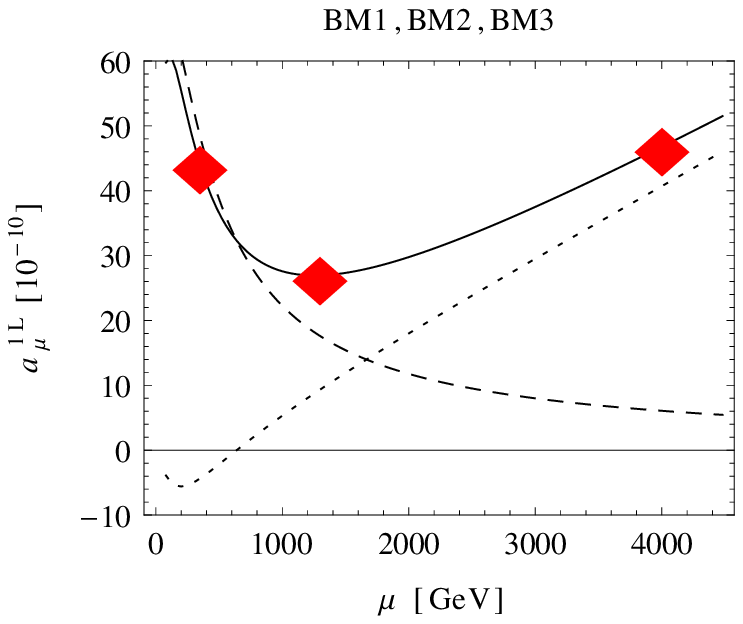}%
\epsfxsize=0.5\textwidth\epsfbox{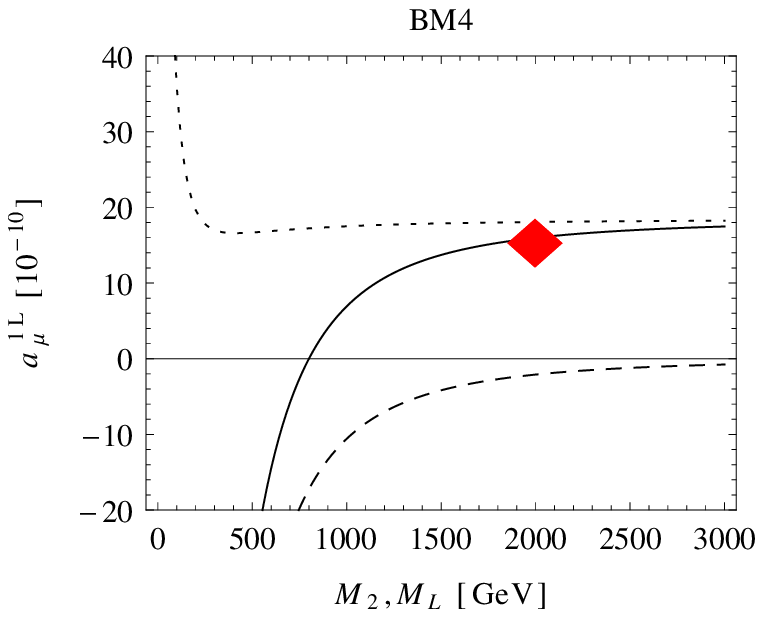}
\caption{\label{fig:oneloopcases}
Numerical results for $\amuSUOL$
as a function of $\mu$ (left) and $M_2=M_L$ (right). The other
parameters are set to the values of BM1--BM3 (left) and BM4 (right). The total one-loop contribution
is drawn solid, the 
chargino/neutralino contributions are drawn dashed/dotted,
respectively. The
benchmark points are indicated as red diamonds. }
\end{center}
\end{figure*}

Before describing the new contributions in more detail we give a brief
account of the theory status of the SUSY prediction for $a_\mu$. The
one-loop contributions in \mbox{Fig.\ \ref{fig:diagrams}}
arise from Feynman diagrams with the exchange
of the SUSY partners of the muon or neutrino, smuon $\tilde{\mu}$ or
sneutrino $\tilde{\nu}_\mu$, and
the SUSY partners of the neutral or charged SM bosons, the neutralinos
or charginos $\tilde{\chi}^{0,\pm}$. They have been computed in the
general MSSM in Ref.\ 
\cite{moroi}, see also \cite{MartinWells,DSreview}.
If all relevant SUSY masses are equal to a common scale $M_{\rm
  SUSY}$, these contributions can be approximated by $13 \times
10^{-10} \,{\mbox{sgn}}(\mu) 
\tan\beta \left({100\ {\rm GeV}  / M_{\rm SUSY}}\right)^2$,
where $\mu$ is the Higgsino mass parameter and $\tan\beta=v_u/v_d$ 
is the ratio of the Higgs doublet vacuum expectation values. This
clearly shows that SUSY can be the origin of the deviation
(\ref{deviation}). However, in the general case, already the one-loop contributions have an
intricate parameter dependence, as shown on an analytic level in
\mbox{Refs.\
  \cite{moroi,MartinWells,DSreview,Cho:2011rk}}. We illustrate this variety of 
possibilities at the one-loop level by \mbox{Fig.\  \ref{fig:oneloopcases}} 
and define representative benchmark parameter points in Tab.\ \ref{BMDefinition}.
\begin{table}
\begin{center}
\scalebox{1}{
\begin{tabular}{l c c c c c c}
					& BM1	& BM2	& BM3	& BM4	\\
\hline\hline
$\mu [\text{GeV}]$			& 350	& 1300	& 4000	& $-160$	\\
$\text{tan}\beta$			& 40	& 40	& 40	& 50	\\
$M_1 [\text{GeV}]$			& 150	& 150	& 150	& 140	\\
$M_2 [\text{GeV}]$			& 300	& 300	& 300	& 2000	\\
$M_{E} [\text{GeV}]$			& 400	& 400	& 400	& 200	\\
$M_{L} [\text{GeV}]$ 			& 400	& 400	& 400	& 2000	\\
$\amuSUOL[ 10^{-10}]$	& 44.02	& 26.95	& 46.78	& 15.98
\end{tabular}
}
\end{center}
\caption{\label{BMDefinition} Definition of the benchmark points.
}
\end{table}
The points are similar to the scenarios studied in Ref.\
\cite{Endo:2013bba}, where it was also shown that such parameter
choices, together with the squark masses defined below, are compatible
with current LHC data.  
We use a notation similar to Ref.\ \cite{DSreview}; $M_{1,2}$ are the
gaugino masses, and the squark and slepton doublets and singlets are
denoted as $M_{Qi}$, $M_{Ui}$, $M_{Di}$, $M_{Li}$, $M_{Ei}$, for each generation
$i\in\{1,2,3\}$. For simplicity we choose
generation-independent masses for the first two
generations, $M_{E1}=M_{E2}\equiv M_{E}$, $M_{L1}=M_{L2}\equiv M_{L}$,
etc., and we set all trilinear soft SUSY breaking $A$ parameters to
zero. 

\mbox{Fig.\ \ref{fig:oneloopcases}}(left) shows the one-loop results $\amuSUOL$ 
as a function of $\mu$ (the remaining parameters are chosen as in BM1--BM3).
There are clearly three characteristic regions for $\mu$, represented
by the three benchmark points BM1--BM3. In the small-$\mu$ region (BM1),
$\mu$ and all other masses are similar. In terms of mass-insertion
diagrams (see e.g.\ Refs.\ \cite{moroi,DSreview}), the dominant
contribution is the diagram with Higgsino--wino and sneutrino
exchange. The result drops with increasing $\mu$ because of the
Higgsino propagator. For intermediate $\mu$ there is a minimum (BM2),
and  the chargino and neutralino contributions become similar. For
large $\mu$ the one-loop results increase linearly in  $\mu$ (BM3).
In this large-$\mu$ region of parameter space all contributions
involving Higgsinos are suppressed,
but the mass-insertion diagram with pure bino exchange and
\mbox{$\tilde{\mu}_R$-$\tilde{\mu}_L$}-transition in the smuon
line is linear in $\mu$ and contributes significantly.

Fig.\ \ref{fig:oneloopcases}(right) shows 
the one-loop results as a function of $M_{L}$ (where $M_2=M_{L}$ and the remaining parameters
are chosen as in BM4). Again, for small $M_{L}$ all masses are similar and the
equal-mass approximation applies. For large $M_{L}$ all contributions
involving $\tilde{\mu}_L$ or a sneutrino are
suppressed, but the mass-insertion diagram with a purely
right-handed smuon 
$\tilde{\mu}_R$ and Higgsino--bino exchange is large. This contribution has the
opposite sign, so $\mu$ is 
chosen negative to allow a positive contribution to $a_\mu$. The
benchmark point BM4 represents this large $M_{L}$-region. The
possibility for such contributions with negative $\mu$ and their
compatibility with dark matter constraints was also stressed recently
in Ref.\ \cite{Grothaus:2012js}.

Our benchmark points are deliberately not optimized to give a
particular value for $a_\mu$ but to represent characteristic regions
of parameter space, for the following reasons. First, the result for $a_\mu$ can
be adjusted easily to the value of Eq.\ (\ref{deviation}) or any
future measurement by tuning the values of $\tan\beta$ and the SUSY
masses, without changing the characteristic of the parameter
point. Second, our later two-loop results will hold in larger regions
of parameter space represented by the benchmark points.

The SUSY two-loop contributions can be grouped into two classes \cite{DSreview}:
In class 2L(a) a pure SUSY loop (of either charginos, neutralinos, or
sfermions) is inserted into a SM-like diagram;
these contributions
have been evaluated exactly in Refs.\ \cite{HSW03,HSW04}. They can be large in certain
regions of parameter space, but they become small as the masses of
SUSY particles and/or heavy Higgs bosons become large. Diagrams of
class 2L(b) involve a loop
correction to a SUSY one-loop diagram. Their results are not completely known, but
leading QED-logarithms \cite{DG98}, the full QED-corrections
\cite{vonWeitershausen:2010zr}, and $(\tan\beta)^2$-enhanced corrections
\cite{Marchetti:2008hw} have been computed. Further 
computations of selected diagrams of classes 2L(a) and 2L(b)  have
been carried out in Refs.\ \cite{Feng1,Feng2,Feng3,FengLM06}. 
All known contributions of class 2L(b) can be significant corrections to
the SUSY one-loop contributions.

We close the section by listing our standard values for the additional
parameters that become relevant at the two-loop level, the additional  squark
and slepton mass parameters  $M_{U,D,Q,U3,D3,Q3}$ and $M_{E3,L3}$ and the
CP-odd Higgs-boson mass  $M_A$.
Where not stated otherwise, we set, like Ref.\
\cite{Endo:2013bba},
\begin{align}
M_{U,D,Q,U3,D3,Q3}&=7\mbox{ TeV},\nonumber\\
M_{E3,L3}&=3\mbox{ TeV},\\
M_A&=1.5\mbox{ TeV}.\nonumber
\end{align}

\section{Fermion/sfermion-loop contributions}

The two-loop fermion/sfermion-loop contributions of
Fig.\ \ref{fig:diagrams} considered in the present Letter belong to the class
2L(b). Their computation represents a big step towards the full
two-loop computation of $a_\mu$ in the MSSM. It is of interest to
consider these contributions separately since they are gauge
invariant and finite by themselves, since their evaluation is
particularly challenging, and since they have a distinctive
phenomenological behaviour.

The fermion/sfermion pairs in the
inner loop run over all quarks and leptons of all three generations
and the associated squarks and sleptons. 
Hence these contributions
introduce a dependence of $a_\mu$ on all sfermion mass parameters
for  the squark and slepton doublets and singlets of all
generations. With the assumption of universality of the first two 
generations stated above,  and since the smuon mass parameters $M_L$
and $M_E$ appear already in the one-loop contributions,  
eight new free mass parameters for the inner loops come into play: $M_U$, 
$M_D$, $M_Q$, $M_{U3}$, $M_{D3}$, $M_{Q3}$, $M_{E3}$, $M_{L3}$.

Technically, the two-loop fermion/sfermion-loop diagrams involve a
higher number of heavy mass scales than all previously considered 
two-loop contributions to $a_\mu$ in the SM and the MSSM. The
SM diagrams involve up to three, the diagrams of Refs.\
\cite{HSW03,HSW04} up to four, the diagrams here up to five different
heavy mass scales.
We have computed the diagrams in two ways --- once by
appropriately extending the standard techniques developed for Refs.\
\cite{HSW03,HSW04}, and once using an iterated one-loop calculation
similar to the simpler cases of Ref.\ \cite{BarrZee}.
A similar class of diagrams with neutralino or gluino exchange has
been considered for electric dipole moments in
Refs. \cite{Yamanaka:2012qn,Yamanaka:2012ia} in an approximation where
Higgsino--gaugino mixing is neglected. In Refs.\
\cite{BarrZee,Yamanaka:2012qn,Yamanaka:2012ia} all two-loop diagrams
were ultraviolet finite, while in our case the diagrams involve
subdivergences and need to be renormalized.

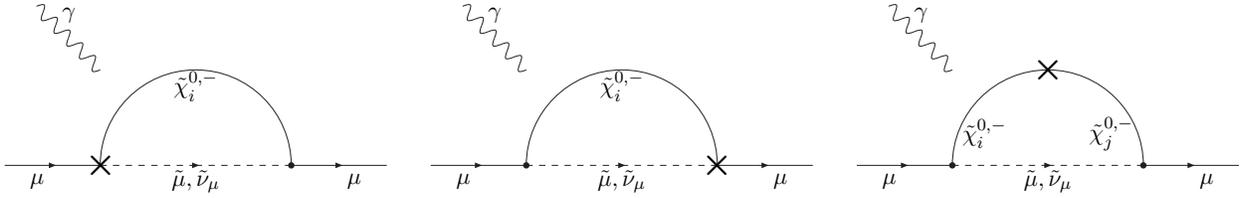
\begin{figure*}
\begin{center}
\scalebox{.6}{\setlength{\unitlength}{1pt}
\begin{picture}(240,100)(-120,0)
\CArc(0,0)(60,0,180)
\ArrowLine(-120,0)(-60,0)
\ArrowLine(60,0)(120,0)
\DashArrowLine(-60,0)(60,0){4}
\Photon(-100,100)(-60,60){4}{5}
\Text(-60,0)[c]{\scalebox{2}{\boldmath{$\times$}}}
\Vertex(60,0){2}
\Text(-80,95)[]{\scalebox{1.35}{$\gamma$}}
\Text(-100,-10)[]{\scalebox{1.35}{$\mu$}}
\Text(100,-10)[]{\scalebox{1.35}{${\mu}$}}
\Text(0,-10)[c]{\scalebox{1.35}{$\tilde{\mu}, \tilde{\nu}_{\mu}$}}
\Text(0,50)[c]{\scalebox{1.35}{$\tilde{\chi} _{i} ^{0,-}$}}
\end{picture}}
\hfill
\scalebox{.6}{\setlength{\unitlength}{1pt}
\begin{picture}(240,100)(-120,0)
\CArc(0,0)(60,0,180)
\ArrowLine(-120,0)(-60,0)
\ArrowLine(60,0)(120,0)
\DashArrowLine(-60,0)(60,0){4}
\Photon(-100,100)(-60,60){4}{5}
\Text(60,0)[c]{\scalebox{2}{\boldmath{$\times$}}}
\Vertex(-60,0){2}
\Text(-80,95)[]{\scalebox{1.35}{$\gamma$}}
\Text(-100,-10)[]{\scalebox{1.35}{$\mu$}}
\Text(100,-10)[]{\scalebox{1.35}{${\mu}$}}
\Text(0,-10)[c]{\scalebox{1.35}{$\tilde{\mu}, \tilde{\nu}_{\mu}$}}
\Text(0,50)[c]{\scalebox{1.35}{$\tilde{\chi} _{i} ^{0,-}$}}
\end{picture}}
\hfill
\scalebox{.6}{\setlength{\unitlength}{1pt}
\begin{picture}(240,100)(-120,0)
\CArc(0,0)(60,0,180)
\ArrowLine(-120,0)(-60,0)
\ArrowLine(60,0)(120,0)
\DashArrowLine(-60,0)(60,0){4}
\Photon(-100,100)(-60,60){4}{5}
\Text(0,60)[c]{\scalebox{2}{\boldmath{$\times$}}}
\Vertex(60,0){2}
\Vertex(-60,0){2}
\Text(-80,95)[]{\scalebox{1.35}{$\gamma$}}
\Text(-100,-10)[]{\scalebox{1.35}{$\mu$}}
\Text(100,-10)[]{\scalebox{1.35}{${\mu}$}}
\Text(0,-10)[c]{\scalebox{1.35}{$\tilde{\mu}, \tilde{\nu}_{\mu}$}}
\Text(-40,20)[]{\scalebox{1.35}{$\tilde{\chi} _{i} ^{0,-}$}}
\Text(40,20)[]{\scalebox{1.35}{$\tilde{\chi} _{j} ^{0,-}$}}
\end{picture}}
\end{center}
\caption{\label{fig:ctdiagrams}
Relevant counterterm diagrams with counterterm insertions at
  the external muon vertex or in the chargino/neutralino self
  energy. The photon can couple to each charged particle.}
\end{figure*}

In addition to the genuine two-loop diagrams we consider the
counterterm diagrams of Fig.\ 
\ref{fig:ctdiagrams}. We use the on-shell renormalization scheme for
SM and SUSY masses as in Refs.\
\cite{Denner93,HKRRSS,tf,Heidi,footnoterenormalization} and the
$\overline{\rm DR}$-scheme for $\tan\beta$ \cite{Freitas:2002um}.
The renormalization constants are computed from diagrams with 
either mixed fermion/sfermion-loops or pure fermion- or
sfermion-loops. Adding two-loop and counterterm diagrams yields the
final finite and well-defined result for the fermion/sfermion-loop
contributions $\amuFSf$.

The full details of the calculation and analytical results will be
presented in a forthcoming 
publication. In the following we will present and discuss the main
properties of the results for $\amuFSf$. Their two most prominent features are:
\begin{itemize}
\item They contain the large and universal
quantities  $\Delta\alpha$ and $\Delta\rho$ from fermion and sfermion
loops, see Sec.\ \ref{sec:universalcorrections}. 
\item They show non-decoupling behaviour if e.g.\ squark masses become
  large, and they contain large logarithms of ratios of squark masses
  over smuon, chargino and neutralino masses, see Sec.\ \ref{sec:logs}. This allows to find
  rather simple semianalytic approximations for these contributions.
\end{itemize}

\subsection{\label{sec:universalcorrections}Large universal corrections}

To discuss the first point we start by 
noting that all SUSY one-loop contributions are
proportional to the fine-structure constant $\alpha=e^2/4\pi$ and some
power of the weak mixing angle $s_W$ or $c_W$. There are several
motivated definitions of these quantities, e.g.\ $\alpha$ can be
defined at zero momentum in the Thomson limit, or as a running
$\alpha(M_Z)$  at the Z-boson mass scale, or $\alpha/s_W^2$ could be
eliminated in favour of the muon decay constant $G_F$. Here,
$\alpha(M_Z)=\alpha/(1-\Delta\alpha(M_Z))$, and the shift
$\Delta\alpha(M_Z)$ is defined via fermion-loop contributions to the
photon vacuum polarization, see Ref.\ 
\cite{HMNT} for a recent evaluation. Using $G_F$ effectively amounts
to using $\alpha(G_F)=\alpha(1+\Delta r)$, where $\Delta r$ summarizes quantum
corrections to muon decay. The leading contributions to $\Delta r$ are
given by $\Delta\alpha(M_Z)$ and the fermion- and sfermion-loop
contributions to the quantity $\Delta\rho$, see Ref.\
\cite{Heinemeyer:2006px} for definitions and a precise MSSM
evaluation. Inserted into $\amuSUOL$,
the differences are numerically sizeable, and this ambiguity is an
inherent source of theory uncertainty of the one-loop
calculation. However, the differences are formally of two-loop order,
and in a full two-loop calculation the differences will be compensated
by corresponding differences in the definitions of the renormalization
constant $\delta e$.

The point is that the differences arise mainly from
$\Delta\alpha(M_Z)$ and $\Delta\rho$ --- and thus from fermion and
sfermion loops. So it is precisely  
our new class of two-loop contributions to $a_\mu$ which eliminates
the one-loop parametrization ambiguity. The full two-loop result is
insensitive to the choice of parametrization, up to subleading
contributions to $\Delta r$ and three-loop effects. In our calculation we choose
to parametrize the one-loop result in terms of $\alpha(M_Z)$, which
leads to the simplest structure of the renormalization constants and
the smallest values of the two-loop contributions.
\begin{table}
\begin{center}
\scalebox{1}{
\begin{tabular}{l c c c c}
				& BM1 		& BM2		& BM3		& BM4		\\
\hline\hline
$\amuSUOL(\alpha)$		& $41.42$ 	& $25.36$	& $44.02$	& $15.03$	\\
$\amuSUOL(\alpha(G_F))$	& $42.92$	& $26.28$	& $45.62$	& $15.58$	\\
$\amuSUOL(\alpha(M_Z))$	& $44.02$	& $26.95$	& $46.78$	& $15.98$	\\
\hline
$\amuSUOL+\amuFSf$	& $45.82$	& $28.16$	& $48.98$	& $16.76$
\end{tabular}
}
\caption{\label{tab:deltarhoresults}
One-loop results for different choices of the 
fine-structure constant $\alpha$ and our two-loop results, which remove
the one-loop ambiguity. The results are given in units of $10^{-10}$.}
\end{center}
\end{table}

Tab.\ \ref{tab:deltarhoresults} shows the numerical impact of the
one-loop parametrization ambiguity compared to our full two-loop
result, for the cases of the benchmark points. The first rows show the
three different one-loop results obtained from the three indicated
definitions of $\alpha$; the  last row shows the sum of the one-loop
result and our new two-loop contributions, consistently parametrized
in terms of $\alpha(M_Z)$. The size of the intervals spanned
by the different one-loop results is $6\%$. The two-loop contributions
are between $4\%$ and $5\%$ of the respective one-loop contributions.
It is not surprising that the two-loop contributions are similarly
large as the size of these one-loop intervals; but it is noteworthy
that in all cases the full two-loop result is significantly {\em
  outside} the one-loop intervals. 
This highlights the importance of the fermion/sfermion-loop
contributions, beyond merely reducing the theory uncertainty.

\begin{figure*}
\begin{center}
\epsfxsize=0.8\textwidth\epsfbox{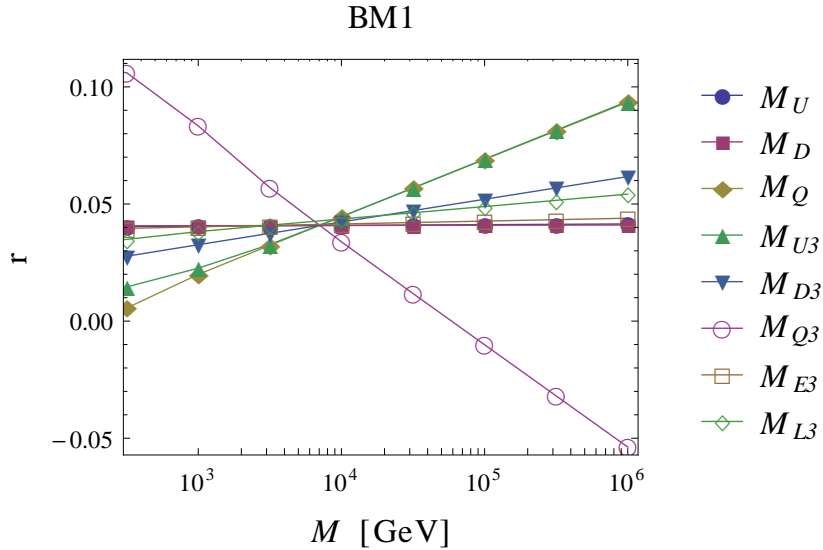}
\end{center}
\caption{\label{fig:BM1plot}
Relative correction $r\equiv\amuFSf/{\amuSUOL}$ from
  fermion/sfermion loops
for benchmark point BM1 as a function of each sfermion mass parameter.
}
\end{figure*}

\subsection{\label{sec:logs}Parameter dependence and non-decoupling behaviour}

Next we focus on the parameter dependence of the new two-loop
contributions. As discussed above, in our analysis there are eight
free mass parameters for the inner loops, in addition to the
parameters already present at the one-loop level. 

Motivated by the LHC results discussed in the introduction, we
allow split spectra and vary all the eight mass parameters
separately over the wide range from 1\ldots1000~TeV. Fig.\
\ref{fig:BM1plot} shows the resulting two-loop corrections for the
case of benchmark point BM1, where one mass parameter is varied at a
time and the others are fixed at their standard values. The
non-decoupling, logarithmic dependence on all the sfermion 
masses is apparent. For this benchmark point and the chosen sfermion
mass range, the two-loop
corrections can have both signs and can be as large as 10\% or $-5$\%
of the one-loop contributions.

More generally, the results for
sufficiently heavy sfermions can be well approximated by leading
logarithms. We obtain the semianalytical expression
\begin{align}
\frac{\amuFSf}{\amuSUOL} \approx
b_0
&+\sum_{\tilde{q}}b_{\tilde{q}}\log\frac{M_{\tilde{q}}}{\mbox{7 TeV}}
\nonumber\\
&+\sum_{\tilde{l}}b_{\tilde{l}}\log\frac{M_{\tilde{l}}}{\mbox{3 TeV}},
\label{LLapproximation}
\end{align}
where the sums extend over the squarks 
\mbox{$\tilde{q}=Q,U,D,Q3,U3,D3$} and the sleptons \mbox{$\tilde{l}=L3,E3$},
respectively. The coefficients $b_0$, $b_{\tilde{q}}$, $b_{\tilde{l}}$
implicitly depend
on the one-loop parameters in a complicated way, but the dependence on
the eight two-loop parameters is made explicit.
We have verified that this approximation is valid for all benchmark
points provided the sfermion masses are above around 5 TeV.
Instead of repeating Fig.\ \ref{fig:BM1plot} for all benchmark points
we provide \mbox{Tab.\ \ref{tab:coefficients}} with all necessary
coefficients.

The physical reason for the non-decoupling behaviour is that the effective
theory obtained by integrating out a heavy sfermion is not
supersymmetric any more. Gaugino and Higgsino interactions can
differ from the corresponding gauge and Yukawa
interactions by matching constants which contain large logarithms of
the heavy sfermion mass. For gaugino interactions these non-decoupling
matching corrections are known as superoblique corrections, and
Refs.\ \cite{Cheng:1997sq,Katz:1998br,Kiyoura:1998yt} have given analytical results for the
coefficients of the large logarithms for several cases. The results
are proportional to the square of the gauge couplings of the heavy
sfermion. Likewise, the matching corrections to the Higgsino
interactions 
(analytical results for heavy squarks can be found in Ref.\
\cite{Chankowski:1989du})
contain large logarithms times the square of the sfermion
Yukawa coupling. This knowledge allows a qualitative understanding of
the results of Fig.\ \ref{fig:BM1plot} and Tab.\
\ref{tab:coefficients}.

The top and bottom Yukawa couplings are the largest couplings of the
inner loop, followed by the tau Yukawa and the SU(2) and U(1) gauge couplings. The
first/second generation Yukawa couplings are 
negligible. In BM1, the Higgsino--wino contribution dominates at the
one-loop level. Hence, the slopes of the $M_{Q3}$, $M_{U3}$
lines in Fig.\ \ref{fig:BM1plot} are largest because of the large
top Yukawa coupling to the Higgsino. The slope of the $M_Q$
line is also large since the left-handed squarks couple to the
wino. The slopes of the $M_{U}$, $M_{D}$ 
lines are particularly small since the 1st/2nd generation singlets
have neither significant Yukawa couplings nor SU(2) gauge
interactions.

%
%
\begin{table}
\begin{center}
\scalebox{1}{
\begin{tabular}{l rrrr}
		& BM1		& BM2		& BM3		& BM4		\\
\hline\hline
$b_0$		& 0.0408	& 0.0446	& 0.0469	& 0.0490	\\
$b_{Q}$		& 0.0106 	& 0.0060	& 0.0014	& $-0.0011$	\\
$b_{U}$		& 0.0001	& 0.0013	& 0.0025	& 0.0031	\\
$b_{D}$		& 0.0000	& 0.0003	& 0.0006	& 0.0008	\\
$b_{Q3}$	& $-0.0190$	&$-0.0106$	&$-0.0019$	& 0.0355	\\
$b_{U3}$	& 0.0107	& 0.0069	& 0.0025	& $-0.0448$	\\
$b_{D3}$	& 0.0042	& 0.0024	& 0.0007	& 0.0075	\\
$b_{L3}$	& 0.0023	& 0.0015	& 0.0007	& 0.0014	\\
$b_{E3}$	& 0.0005	& 0.0008	& 0.0010	& 0.0018
\end{tabular}
}
\end{center}
\caption{\label{tab:coefficients}
Coefficients of Eq.\ \eqref{LLapproximation}
for benchmark points BM1--BM4.}
\end{table}

\begin{figure*}
\epsfxsize=0.49\textwidth\epsfbox{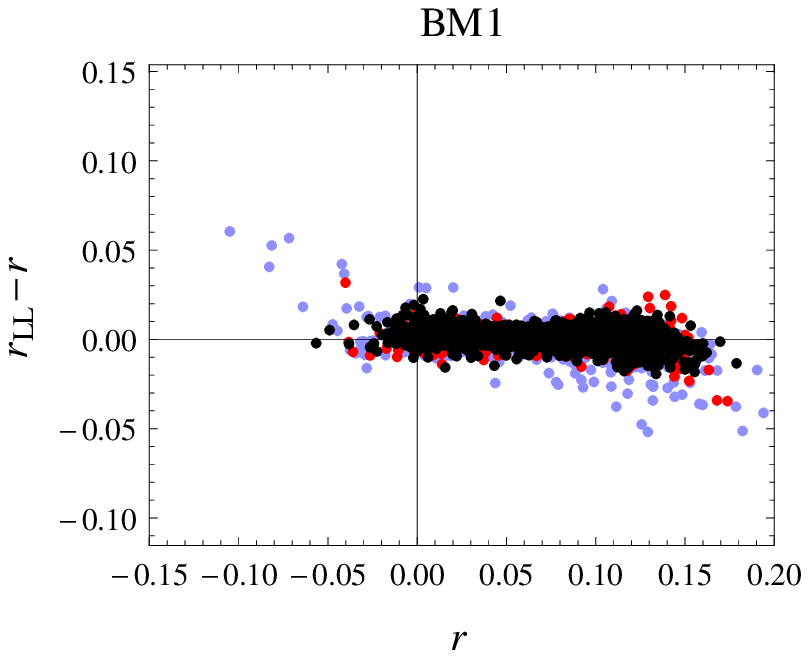}
\epsfxsize=0.49\textwidth\epsfbox{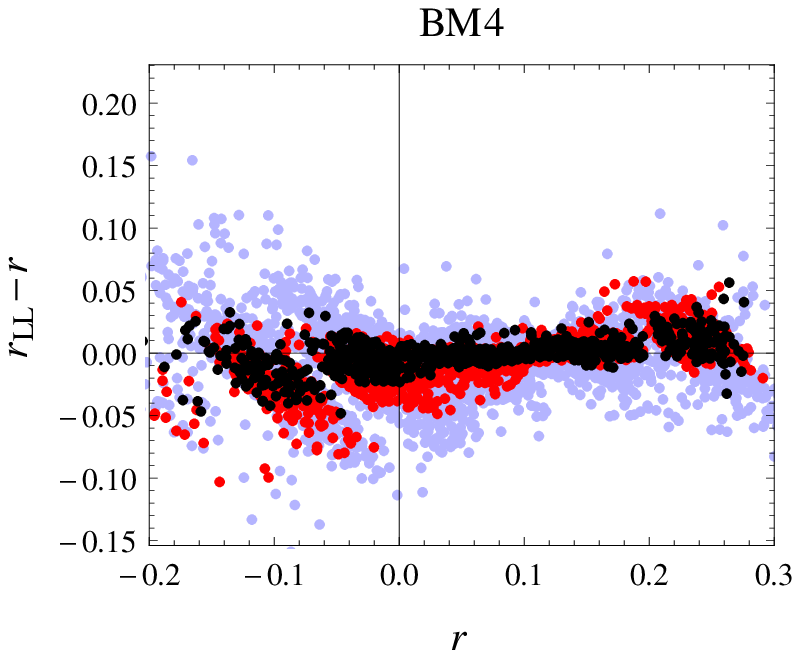}
\caption{\label{fig:scatter}
The exact result for $r\equiv{\amuFSf}/{\amuSUOL}$
compared with the approximation $r_{\rm LL}$ given by the r.h.s.\ of Eq.\
  (\ref{LLapproximation}) together with the coefficients in Tab.\
  \ref{tab:coefficients}. The mass parameters are chosen randomly
  around the benchmark points BM1 and BM4, with the ranges given in
  the text.  }  
\end{figure*}

As the parameters are changed from BM1 to BM2 and BM3, the Higgsino--wino contribution becomes
less, the pure bino contribution more important. 
Clearly, the non-decoupling
logarithms to the bino contributions are controlled by the small U(1)
gauge couplings and the hypercharges. We have verified that
for a parameter point where the bino contribution dominates
sufficiently strongly, the coefficients of the
logarithms indeed become generation-independent and proportional to the
squared hypercharge of the respective sfermion. For our point BM3,
only the coefficients $b_{U}$, $b_D$, $b_{E3}$ have
this behaviour. The other coefficients also receive
non-negligible corrections from the Higgsino--wino contributions,
which are about 7 times smaller than the bino contribution but involve
much larger corrections similar to the case
of BM1. Nevertheless, this explains the overall decrease of the slopes
for BM2 and BM3.

The case of BM4 is similar to the one of BM1, except that at the
one-loop level the Higgsino--bino contribution dominates, instead of
the Higgsino--wino contribution. The relative corrections from
top-Yukawa enhanced contributions are larger for BM4, and therefore
the slopes for $\log M_{Q3,U3}$ are even larger for BM4 than for BM1.

The validity of the approximation (\ref{LLapproximation}) and the
coefficients in Tab.\ \ref{tab:coefficients} goes far beyond the pure benchmark
points, because the benchmark points are
representative for larger regions of parameter space with
characteristic properties. Whenever we choose a parameter point
similar to one of the benchmark points, we can still use Eq.\
(\ref{LLapproximation}) with the coefficients given in the table. The
quality of the approximation is quantified in Fig.\ \ref{fig:scatter}.
It shows the difference of the exact result and the approximation by
Eq.\ (\ref{LLapproximation}) for randomly chosen parameter points
scattered around the benchmark points.
The parameter ranges for the
light blue points are given in Tab.\ \ref{Lightconstraints}.
\begin{table}
\begin{center}
\scalebox{1}{
\begin{tabular}{l cl c l}
					& BM1	 &	& BM4&	\\
\hline\hline
$\mu [\text{GeV}]$			& [100,200]
&& [-200,-100]
\\
$M_1 [\text{GeV}]$			& [100,200]
&& [100,200]& \\
$M_2 [\text{GeV}]$
& [200,400]
&
& [1000,3000] &
\\
$M_{E} [\text{GeV}]$			& [200,500] &
	 	& [100,300]
                \\
$M_{L} [\text{GeV}]$ 			& [200,500]
&& [1000,3000] \\
\end{tabular}
}
\end{center}
\caption{\label{Lightconstraints} Scan intervals for the least
  restrictive light blue parameter regions.} 
\end{table}
We also impose the constraints 
  $\amuSUOL\ge5\times10^{-10}$ and  $|M_2-\mu |\ge5$~GeV to avoid
  artificially large effects due to accidental cancellations.
 The other eight sfermion-mass parameters 
  are varied in the range $[10^3,10^6]$ GeV.
For the red and black points the further constraints given in Tab.\
\ref{RedBlackconstraints}  are
successively applied.
\begin{table}
\begin{center}
\begin{tabular}{l cl c l}
					& BM1	 &	& BM4&	\\
\hline\hline
red & $M_{E,L}\le400$ &\quad\null&
$\begin{array}{rl}&M_2,M_L\ge2000,\\
& M_1\le|\mu|+40\\[1ex]
\end{array}
$\\[1ex]
black & $M_{E,L}\ge250$ && $M_1\le|\mu|-10$
\end{tabular}
\end{center}
\caption{\label{RedBlackconstraints} Additional parameter constraints
  for the red and black parameter regions (in GeV).}
\end{table}
These further constraints strengthen the equal-mass characteristic of
the BM1 region and the decoupling of the wino and $\tilde{\mu}_L$ in
the BM4 region.

In the most restrictive black region, the parameters can fluctuate within a
factor $\sim 1.5$ around the benchmark points. In this region, the
approximation is generally good. For the red and light blue
parameter regions, the 
approximation becomes gradually worse, but even in the largest
region, the approximation works well for the
majority of parameter points. We only show the results for
BM1 and BM4; for the other benchmark points the corrections are
smaller and the approximation works even better.

The figure also shows the generally possible magnitude of
the fermion/sfermion-loop corrections. Already within the most
restrictive considered parameter
regions, the corrections are up
to 15\% and 30\% of the respective one-loop result for BM1, BM4,
respectively.

\subsection{Comparison with other MSSM two-loop contributions}

\begin{figure*}
\begin{center}
\end{center}
\begin{center}
\epsfbox{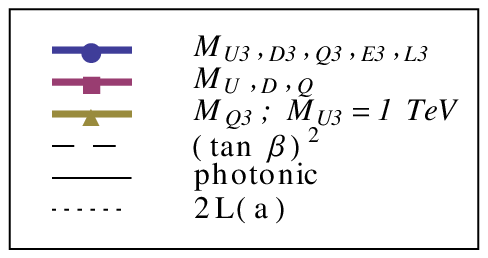}
\\
\epsfxsize=0.49\textwidth\epsfbox{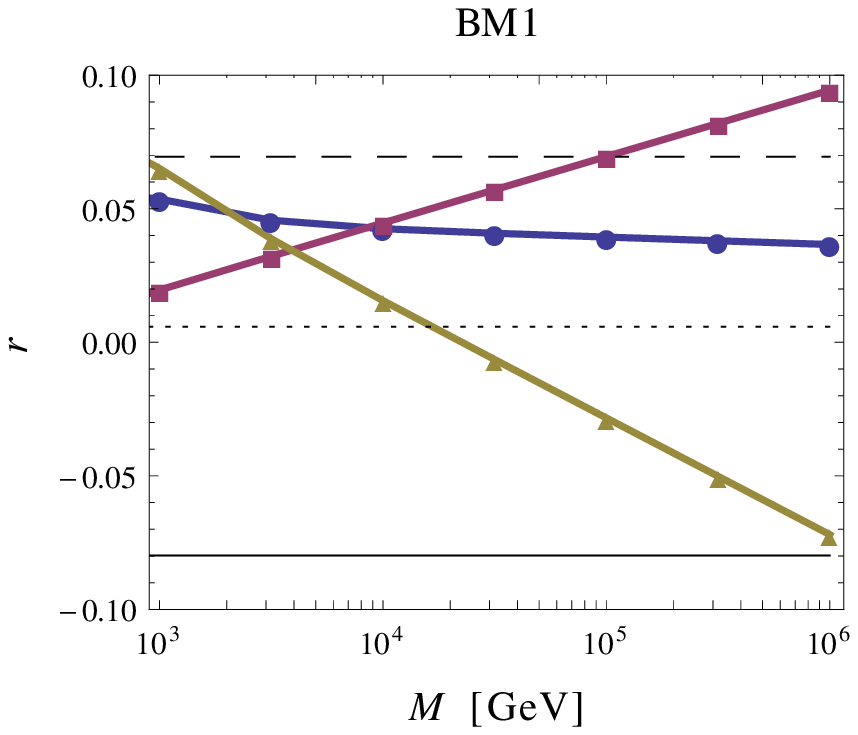}
\epsfxsize=0.49\textwidth\epsfbox{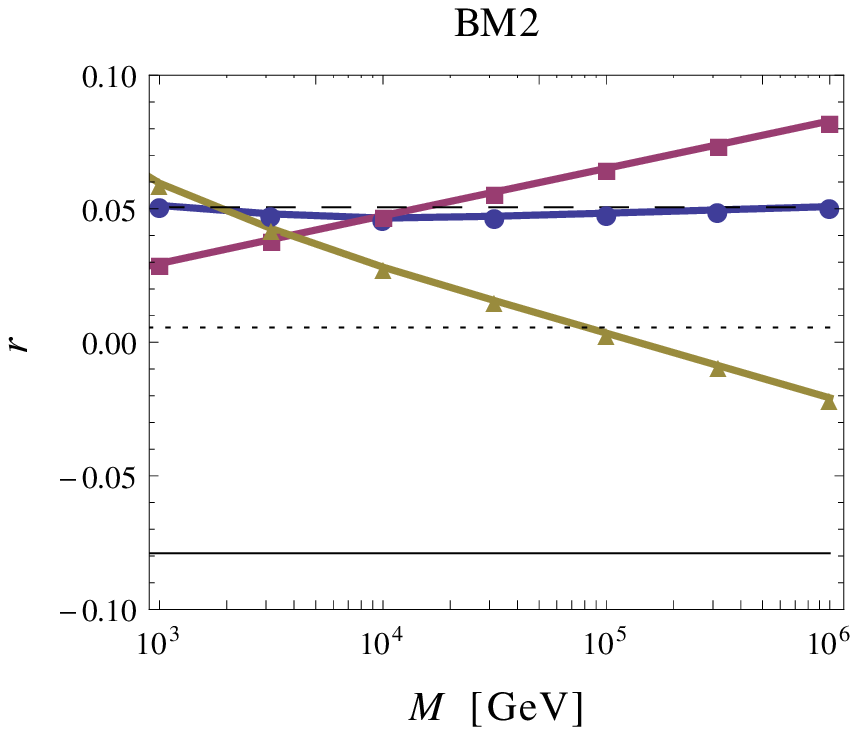}
\\
\epsfxsize=0.49\textwidth\epsfbox{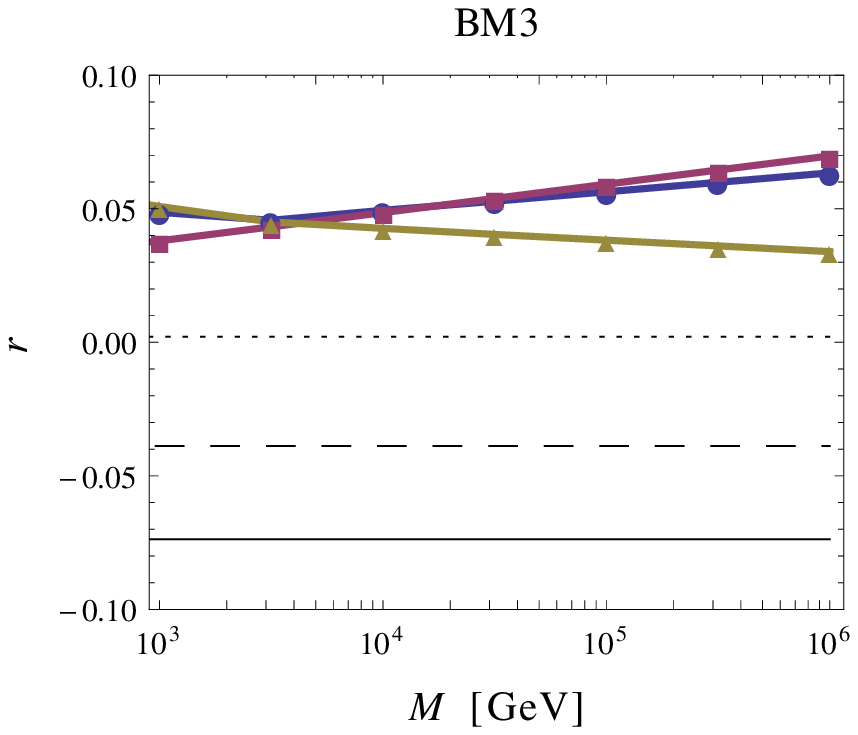}
\epsfxsize=0.49\textwidth\epsfbox{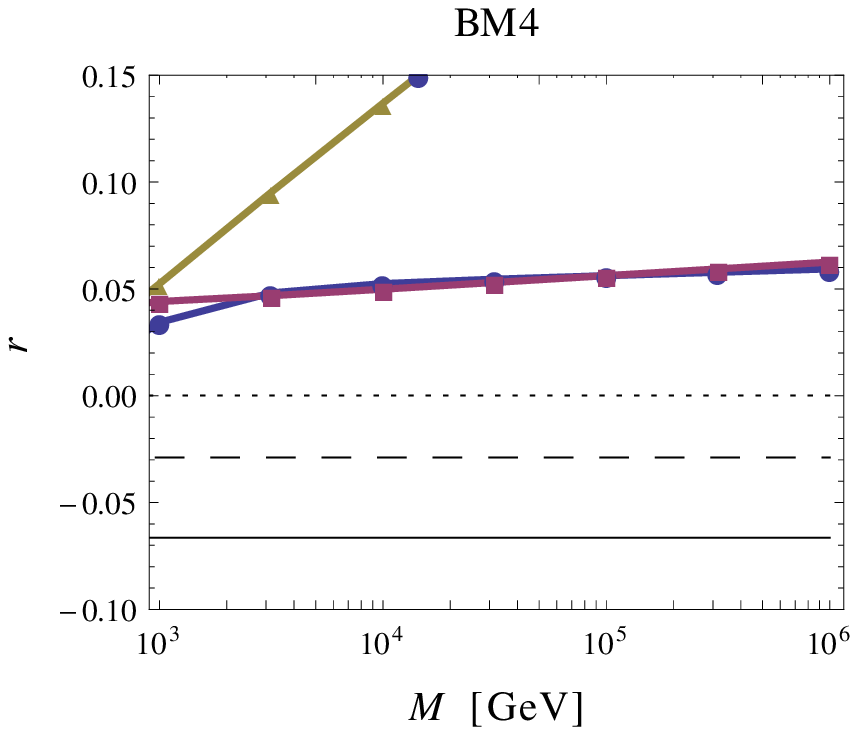}
\end{center}
\caption{\label{fig:twoloopSquarkMasses}
Results for the ratios  $r\equiv{a_\mu^{\rm 2L}}/{\amuSUOL}$
  for the known MSSM two-loop contributions to $a_\mu$,  for all
  benchmark points.  The thick, coloured lines show 
  the new fermion/sfermion-loop contributions for the combinations
of sfermion masses indicated in the legend. The
thin lines show the previously known $(\tan\beta)^2$ (dashed),
photonic (solid), and 2L(a) (dotted) contributions. The  sfermion mass
dependence of the 2L(a) contributions is negligible and invisible in
the plots.}
\end{figure*}

Finally we summarize the behaviour of the fermion/sfermion-loop contributions and
compare it with all previously known two-loop contributions to $a_\mu$
in the MSSM. Fig.\ \ref{fig:twoloopSquarkMasses} shows the results for the
benchmark points as
functions of various motivated combinations of 
sfermion masses: either of a common 
third generation sfermion mass $M_{U3,D3,Q3,E3,L3}\equiv M$, or of a
universal squark mass $M_{U,D,Q}\equiv M$, or, as an example with particularly
large corrections, purely as a function of
$M_{Q3}$ with $M_{U3}$ fixed to 1~TeV. Each time, the non-varied
sfermion masses are kept at their standard values.

As is well known, the photonic contributions 
\cite{vonWeitershausen:2010zr} and the $(\tan\beta)^2$-enhanced 
contributions \cite{Marchetti:2008hw} are both large. The
photonic contributions are around $-8\%$ for the benchmark points BM1--BM3
and around $-7\%$ for BM4 due to its smaller mass scales. The
$(\tan\beta)^2$-enhanced contributions have a non-trivial parameter
dependence, in particular their sign changes when going from BM1 to
either BM3 or BM4. Their magnitude is up to $7\%$ in our examples.
Because of the
heavy Higgs-boson mass the contributions of class 2L(a)  are
small. Even though they have a dependence on the sfermion masses, due
to decoupling in these contributions the dependence is invisible in
the plots.

The new fermion/sfermion-loop contributions are always in the
few-percent range and can even become the largest two-loop
contributions. For the cases of universal squark mass and of common
third-generation sfermion masses we find between 5\% and 10\%
corrections. For the considered case with larger mass splittings,
$M_{Q3}\gg M_{U3}$, the corrections are even more significant, and their
parameter dependence is very strong. There can
be negative corrections for BM1 and BM2. For BM4, the corrections are
positive and around 15\% for squark masses up to 20~TeV; outside the
plot, the
corrections grow beyond 30\% for squark masses up to 1000~TeV.

\section{Conclusions}

Large contributions to $a_\mu$ are
possible in  a variety of qualitatively different parameter regions
of the MSSM. To illustrate this we defined benchmark parameter points
which represent these regions with large $a_\mu$ and which are compatible
with limits from the LHC. 

We then presented results of a new calculation of
fermion/sfermion-loop contributions to $a_\mu$ in the MSSM. These will be
important for a correct interpretation of current or future $a_\mu$
measurements within the MSSM and for drawing precise conclusions on
preferred parameter regions. The new corrections introduce a
dependence of $a_\mu$ on all sfermion masses beyond the smuon masses,
and they can be surprisingly large. For
moderate sfermion masses they are typically around 4\%--5\% but can
also reach up to 10\%, as  Fig.\ \ref{fig:BM1plot} exemplifies. If the
additional sfermion masses are in the 
multi-TeV range, the corrections grow logarithmically similar to the
so-called superoblique corrections; in our examples
up to 30\% corrections are possible particularly in scenarios
with large splitting between left- and right-handed stop masses.

Semianalytical expressions are provided which can be easily
evaluated in practical applications to obtain good estimates of the
new contributions. Together with the formulas provided in Refs.\
\cite{Marchetti:2008hw,vonWeitershausen:2010zr}, this allows a compact
implementation of a good approximation of the MSSM two-loop
contributions to $a_\mu$.

Taking into account the fermion/sfermion-loop contributions removes
the ambiguity from parametrizing the one-loop contributions either in
terms of $\alpha$, $\alpha(M_Z)$, or $G_F$. The full result
including the two-loop corrections is typically outside the interval
spanned by the differently parametrized one-loop results, highlighting
the importance of the new corrections.

Finally, we have illustrated the non-trivial parameter dependence of
the new
fermion/sfermion-loop corrections and  the previously known
two-loop contributions. Both the 
fermion/sfermion-loop and
the $(\tan\beta)^2$
corrections can be positive or negative, and either of them can be
larger in magnitude than the photonic contributions. Each of these new
and previously known two-loop corrections can be larger than the
future experimental uncertainty.

The present Letter focuses on the most prominent features of
the fermion/sfermion-loop corrections. A discussion of the deviations
from the leading logarithmic behaviour and the influence of squark
mixing, together with the full details of the calculation and
analytical results, will be presented in a forthcoming publication. 

\subsection*{Acknowledgements}
We acknowledge financial support by the German Research Foundation DFG through
Grant No. STO876/1-1, by DAAD and by CNPq.
HF thanks TU Dresden and IKTP for their hospitality.

\end{document}